\documentstyle[12pt]{article}                    
\newfont{\ffont}{msym10}                        
\newcommand{\beq}{\begin{equation}}             
\newcommand{\eeq}{\end{equation}}               
\newcommand{\bqry}{\begin{eqnarray}}            
\newcommand{\eqry}{\end{eqnarray}}              
\newcommand{\bqryn}{\begin{eqnarray*}}          
\newcommand{\eqryn}{\end{eqnarray*}}            
\newcommand{\preprint}[1]{\begin{table}[t]      
            \begin{flushright}                  
            \begin{large}{#1}\end{large}        
            \end{flushright}                    
            \end{table}}                        
\newcommand{\PD}[2]                             
    {\frac{\partial^{#2}}{\partial #1^{#2}}}    
\begin{document}
\preprint{LA-UR-96-2415 \\ IASSNS-HEP-96/81}
\title{Gell-Mann--Okubo Mass Formula \\ for $SU(4)$ Meson Hexadecuplet}
\author{\\ L. Burakovsky\thanks{Bitnet: BURAKOV@QCD.LANL.GOV} \
\\  \\  Theoretical Division, T-8 \\  Los Alamos National  
Laboratory \\ Los
Alamos NM 87545, USA \\  \\  and  \\  \\
L.P. Horwitz\thanks{Bitnet: HORWITZ@SNS.IAS.EDU. On sabbatical leave from
School of Physics and Astronomy, Tel Aviv University, Ramat Aviv, Israel.
Also at Department of Physics, Bar-Ilan University, Ramat-Gan,  
Israel  } \
\\  \\ School of Natural Sciences \\ Institute for Advanced Study  
\\ Princeton
NJ 08540, USA \\}
\date{ }
\maketitle
\begin{abstract}
Using a linear mass spectrum of an $SU(4)$ meson hexadecuplet, we derive 
the Gell-Mann--Okubo mass formula for the charmed mesons, in good  
agreement
with experiment. Possible generalization of this method to a higher  
symmetry
group is briefly discussed.
\end{abstract}

\bigskip
\section*{  }
The hadronic mass spectrum is an essential ingredient in theoretical 
investigations of the physics of strong interactions. It is well  
known that
the correct thermodynamic description of hot hadronic matter requires 
consideration of higher mass excited states, the resonances, whose
contribution becomes essential at temperatures $\sim O(100$ MeV)
\cite{Shu,Leut}. The method for taking into account these
resonances was suggested by Belenky and Landau \cite{BL} as considering 
unstable particles on an equal footing with the stable ones in the
thermodynamic quantities; e.g., the formulas for the pressure and energy 
density in a resonance gas read\footnote{For simplicity, we neglect the 
chemical potential and approximate the particle statistics by the
Maxwell-Boltzmann one.}
\beq
p=\sum _ip_i=\sum _ig_i\frac{m_i^2T^2}{2\pi  
^2}K_2\left(\frac{m_i}{T}\right),
\eeq
\beq
\rho =\sum _i\rho _i,\;\;\;\rho _i=T\frac{dp_i}{dT}-p_i,
\eeq
where $g_i$ are the corresponding degeneracies ($J$ and $I$ are spin and
isospin, respectively), $$g_i=\frac{\pi ^4}{90}\times \left[
\begin{array}{ll}
(2J_i+1)(2I_i+1) & {\rm for\;non-strange\;mesons} \\
4(2J_i+1) & {\rm for\;strange}\;(K)\;{\rm mesons} \\
2(2J_i+1)(2I_i+1)\times 7/8 & {\rm for\;baryons}
\end{array} \right. $$
These expressions may be rewritten with the help of a {\it  
resonance spectrum,}
\beq
p=\int _{m_1}^{m_2}dm\;\tau (m)p(m),\;\;\;p(m)\equiv  
\frac{m^2T^2}{2\pi ^2}
K_2\left(\frac{m}{T}\right),
\eeq
\beq
\rho =\int _{m_1}^{m_2}dm\;\tau (m)\rho (m),\;\;\;\rho (m)\equiv
T\frac{dp(m)}{dT}-p(m),
\eeq
normalized as
\beq
\int _{m_1}^{m_2}dm\;\tau (m)=\sum _ig_i,
\eeq
where $m_1$ and $m_2$ are the masses of the lightest and heaviest  
species,
respectively, entering the formulas (1),(2).

In both the statistical bootstrap model \cite{Hag,Fra} and the dual  
resonance
model \cite{FV}, a resonance spectrum takes on the form
\beq
\tau (m)\sim m^a\;e^{m/T_0},
\eeq
where $a$ and $T_0$ are constants. The treatment of a hadronic  
resonance gas
by means of the spectrum (6) leads to a singularity in the thermodynamic 
functions at $T=T_0$ \cite{Hag,Fra} and, in particular, to an  
infinite number
of the effective degrees of freedom in the hadron phase, thus hindering a
transition to the quark-gluon phase. Moreover, as shown by Fowler  
and Weiner
\cite{FW}, an exponential mass spectrum of the form (6) is  
incompatible with
the existence of the quark-gluon phase: in order that a phase  
transition from
the hadron phase to the quark-gluon phase be possible, the hadronic  
spectrum
cannot grow with $m$ faster than a power.

In our previous work \cite{spectrum} we considered a model for a  
transition
from a phase of strongly interacting hadron constituents, described by a 
manifestly covariant relativistic statistical mechanics which  
turned out to be
a reliable framework in the description of realistic physical systems 
\cite{mancov}, to the hadron phase described by a resonance  
spectrum, Eqs.
(3),(4). An example of such a transition is what might be  
considered to be a
relativistic high temperature Bose-Einstein condensation studied by  
the authors
in ref. \cite{cond}, which corresponds, in the way suggested by  
Haber and
Weldon \cite{HW}, to spontaneous flavor symmetry breakdown,  
$SU(3)_F\rightarrow
SU(2)_I\times U(1)_Y,$ upon which hadronic multiplets are formed,  
with the
masses obeying the Gell-Mann--Okubo formulas \cite{GMO}
\beq
m^\ell =a+bY+c\left[ \frac{Y^2}{4}-I(I+1)\right];
\eeq
here $I$ and $Y$ are the isospin and hypercharge, respectively,  
$\ell $ is 2
for mesons and 1 for baryons, and $a,b,c$ are independent of $I$  
and $Y$ but,
in general, depend on $(p,q),$ where $(p,q)$ is any irreducible  
representation
of $SU(3).$ Then only the assumption on the overall degeneracy  
being conserved
during the transition is required to lead to the unique form of a  
resonance
spectrum in the hadron phase:
\beq
\tau (m)=Cm,\;\;\;C={\rm const}.
\eeq
Zhirov and Shuryak \cite{ZS} have found the same result on  
phenomenological
grounds. As shown in ref. \cite{ZS}, the spectrum (8), used in the  
formulas (3),(4) (with the upper limit of integration infinity), leads to
the equation of state $p,\rho \sim T^6,$ $p=\rho /5,$ called by  
Shuryak the
``realistic'' equation of state for hot hadronic matter \cite{Shu},  
which has
some experimental support. Zhirov and Shuryak \cite{ZS} have calculated 
the velocity of sound, $c_s^2\equiv dp/d\rho =c_s^2(T),$ with $p$  
and $\rho $
defined in Eqs. (1),(2), and found that $c_s^2(T)$ at first  
increases with $T$
very quickly and then saturates at the value of $c_s^2\simeq 1/3$  
if only the
pions are taken into account, and at $c_s^2\simeq 1/5$ if  
resonances up to
$M\sim 1.7$ GeV are included.

We have checked the coincidence of the results given by the linear  
spectrum (8)
with those obtained directly from Eq. (1) for the actual hadronic  
species with
the corresponding degeneracies, for all well-established
multiplets, both mesonic and baryonic,
%
%
%
%
$f_2^{'}(1525),$
%
%
%
%
%
and found it excellent \cite{spectrum}. Therefore, the theoretical  
conclusion
that a linear spectrum is the actual spectrum in the description of  
individual
hadronic multiplets finds its experimental confirmation as well. In  
our recent
paper \cite{enigmas} we applied a linear spectrum to the problem of 
establishing the correct $q\bar{q}$ assignment for the problematic meson 
nonets, like the scalar, axial-vector and tensor ones, and  
separating out
non-$q\bar{q}$ mesons.

The easiest way to see that a linear spectrum corresponds to the actual 
spectrum of a meson nonet is as follows\footnote{For a baryon  
multiplet, it is
more difficult to show that the mass spectrum is linear, since the
Gell-Mann--Okubo formulas are linear in mass for baryons. More detailed 
discussion is given in \cite{spectrum}.}. Let us calculate the  
average mass
squared for a spin-$s$ nonet:
\beq
\langle m^2\rangle _9\equiv \frac{\sum  
_ig_i\;m_i^2}{\sum_ig_i}=\frac{3m_1^2+
4m_{1/2}^2+m_{0^{'}}^2+m_{0^{''}}^2}{9},
\eeq
where $m_1,\;m_{1/2},\;m_0,\;m_{0^{'}}$ are the masses of  
isovector, isospinor,
and two isoscalar states, respectively, and the spin degeneracy, $2s+1,$ 
cancels out. In general, the isoscalar states\footnote{The $\omega  
_{0^{'}}$
is a mostly octet isoscalar.} $\omega _{0^{'}}$ and $\omega  
_{0^{''}}$ are the
octet $\omega _8$ and singlet $\omega _0$ mixed states because of  
$SU(3)$
breaking, $$\omega _{0^{'}}=\omega _8\cos \theta _M- \omega _0\sin  
\theta _M,$$
$$\omega _{0^{''}}=\omega _8\sin \theta _M+ \omega _0\cos \theta  
_M,$$ where
$\theta _M$ is a mixing angle. Assuming that the matrix element of the 
Hamiltonian between the states yields a mass squared, i.e., $m_{0^{'}}^2=
\langle \omega _{0^{'}}|H|\omega _{0^{'}}\rangle $ etc., one  
obtains from the
above relations \cite{Per},
\beq
m_{0^{'}}^2=m_8^2\cos ^2\theta _M+m_0^2\sin ^2\theta  
_M-2m_{08}^2\sin \theta _M
\cos \theta _M,
\eeq
\beq
m_{0^{''}}^2=m_8^2\sin ^2\theta _M+m_0^2\cos ^2\theta  
_M+2m_{08}^2\sin \theta
_M\cos \theta _M.
\eeq
Since $\omega _{0^{'}}$ and $\omega _{0^{''}}$ are orthogonal, one  
has further
\beq
m_{0^{'}0^{''}}^2=0=(m_8^2-m_0^2)\sin \theta _M\cos \theta _M+  
m_{08}^2(\cos ^2
\theta _M-\sin ^2\theta _M).
\eeq
Eliminating $m_0$ and $m_{08}$ from (10)-(12) yields
\beq
\tan ^2\theta _M=\frac{m_8^2-m_{0^{'}}^2}{m_{0^{''}}^2-m_8^2}.
\eeq
It also follows from (10),(11) that, independent of $\theta _M,$  
$m_{0^{'}}^2+
m_{0^{''}}^2=m_8^2+m_0^2,$ and therefore, Eq. (9) may be rewritten as    
\beq
\langle m^2\rangle _9=\frac{3m_1^2+4m_{1/2}^2+m_8^2+m_0^2}{9}.
\eeq
For the octet, $(3\;m_1,\;4\;m_{1/2},\;1\;m_8),$ the  
Gell-Mann--Okubo formula
(as follows from (7)) is
\beq
4m_{1/2}^2=3m_8^2+m_1^2.
\eeq
Therefore, the average mass squared for the octet is
\beq
\langle m^2\rangle  
_8=\frac{3m_1^2+4m_{1/2}^2+m_8^2}{8}=\frac{m_1^2+m_8^2}{2},
\eeq
where Eq. (15) was used. In the exact $SU(3)$ limit where the $u,$  
$d$ and $s$
quarks have equal masses, all the squared masses of the nonet  
states are equal
as well. Since in this limit all the squared masses of the octet  
states are
equal to the average mass squared of the octet\footnote{In a manifestly 
covariant theory, this holds since a total mass squared is rigorously 
conserved. In the standard framework, for pseudoscalar mesons, this  
is easily
seen by using the lowest order relations \cite{GMOR} $m_1^2\equiv  
m_\pi ^2=
2mB,$ $m_{1/2}^2\equiv m_K^2=(m+m_s)B,$ where $m=(m_u+m_d)/2,$ and  
$B$ is
related to the quark condensate. Therefore, it follows from  
(15),(16) that
$m_8^2=2/3\;(2m_s+m)B,$ $\langle m^2\rangle  
_8=2/3\;(m_s+2m)B=2/3\;(m_u+m_d+
m_s)B.$ In the exact $SU(3)$ limit, $m_u=m_d=m_s=\bar{m},$ and  
hence $m_1^2=
m_{1/2}^2=m_8^2=\langle m^2\rangle _8=2\bar{m}B.$ For higher mass  
mesons,
since the states with equal isospin (and alternating parity) lie on  
linear
Regge trajectories, one may expect the relations of the form  
$(c=C/B)$ $m_1^2=
2mB+C=(2m+c)B,$ $m_{1/2}^2=(m+m_s)B+C=(m+m_s+c)B,$  
$m_8^2=2/3\;(2m_s+m)B+C=
2/3\;(2m_s+m+3/2\;c)B,$ consistent with the Gell-Mann--Okubo  
formula (15),
leading to $m_1^2=m_{1/2}^2=m_8^2=\langle m^2\rangle  
_8=2\bar{m}B+C$ in the
$SU(3)$ limit $m_u=m_d=m_s=\bar{m}.$ For vector mesons, such a  
relation was
obtained by Bal\'{a}zs in the flux-tube fragmentation approach to a  
low-mass
hadronic spectrum \cite{Bal}, $m_\rho ^2=m_\pi ^2+1/2\alpha ^{'},$  
with $\alpha
^{'}$ being a universal Regge slope, in good agreement with the  
experiment.},
Eq. (16), the mass of the singlet should have the same  
value,\footnote{It is
also seen from the relations of a previous footnote: since the total mass
squared of a nonet is proportional to the total mass of quarks the nonet 
members are made of, $\sum _ig_i\;m_i^2=(12m+6m_s)B+9C,$ it follows  
from the
above expressions for $m_1^2,$ $m_{1/2}^2$ and $m_8^2$ that  
$m_0^2=2/3\;(2m+m_
s)B+C=\langle m^2\rangle _8=\langle m^2\rangle _9.$} i.e.,
\beq
m_0^2=\frac{m_1^2+m_8^2}{2}.
\eeq
With Eq. (15), it then follows from (17) that $$m_0^2+m_8^2=2m_{1/2}^2,$$
which reduces, through $m_0^2+m_8^2=m_{0^{'}}^2+m_{0^{''}}^2,$ to
\beq
m_{0^{'}}^2+m_{0^{''}}^2=2m_{1/2}^2,
\eeq
which is an extra Gell-Mann--Okubo mass relation for a nonet. We  
have checked
this relation in a separate paper \cite{linear}, and found that with the 
experimentally available meson masses, the relative error in the  
values on the
l.h.s. and r.h.s. of Eq. (18) does not exceed 3\% for all  
well-established
nonets (except for the pseudoscalar nonet for which Eq. (18) does  
not hold,
perhaps because the $\eta _0$ develops a large dynamical mass due  
to axial
$U(1)$ symmetry breakdown before it mixes with the $\eta _8$ to form the 
physical $\eta $ and $\eta ^{'}$ states). For a singlet-octet  
mixing close to
``ideal'' one, $\tan \theta _M\simeq 1/\sqrt{2};$ it then follows  
from (13)
that $$2m_{0^{'}}^2+m_{0^{''}}^2\simeq 3m_8^2,$$ which reduces, through 
(15),(18), to
\beq
m_{0^{''}}\simeq m_1.
\eeq
Now it follows clearly that the ground states of all well-established 
nonets\footnote{This is also true for $q\bar{q}$ assignment of the  
scalar
meson nonet suggested by the authors in ref. \cite{enigmas}.}  
(except for the
pseudoscalar one) are almost mass degenerate pairs, like $(\rho
,\omega ).$\footnote{It follows from the relations of footnote 4  
that, in the
close-to-ideal mixing case, $m_{0^{'}}^2\simeq 2m_sB+C$ and $m_{0^{''}}^
2\simeq 2mB+C=m_1^2.$} In the close-to-ideal mixing case, Eq. (18)  
may be
rewritten, with the help of (19), as
\beq
m_1^2+m_{0^{'}}^2\simeq 2m_{1/2}^2.
\eeq
This relation for pseudoscalar and vector mesons with the ground  
states being
the mass degenerate pairs $(\pi ,\eta _0)$ and $(\rho ,\omega ),$  
respectively,
was previously obtained by Bal\'{a}zs and Nicolescu using the
dual-topological-unitarization approach to the confinement region  
of hadronic
physics (Eq. (21) of ref. \cite{BN}). With (16) and (17), Eq. (10)  
finally
reduces to
\beq
\langle m^2\rangle _9=\frac{m_1^2+m_8^2}{2},
\eeq
which, of course, coincides with both, $\langle m^2\rangle _8$ in  
(16) and
$m_0^2$ in (17), which is the property of the $SU(3)$ limit (or the 
conservation of a total mass squared in a manifestly covariant theory). 

For the actual mass spectrum of the nonet, the average mass squared  
(9) may be
represented in the form\footnote{Since $m_s>m,$ $m_1<m_{1/2}<m_8,$  
as seen in
the relations of footnote 4. Moreover, $m_1<m_0<m_8,$ and therefore, 
the range of integration in Eq. (22) is $(m_1,\;m_8).$}
\beq
\langle m^2\rangle _9=\frac{\int _{m_1}^{m_8}dm\;\tau (m)\;m^2}{\int
_{m_1}^{m_8}dm\;\tau (m)},
\eeq
and one sees that the only choice for $\tau (m)$ leading to the  
relation (21)
is $\tau (m)=Cm,$ $C={\rm const.}$ Indeed, in this case $$\langle  
m^2\rangle
_9=\frac{\int _{m_1}^{m_8}dm\;m^3}{\int  
_{m_1}^{m_8}dm\;m}=\frac{(m_8^4-m_1^
4)/4}{(m_8^2-m_1^2)/2}=\frac{m_1^2+m_8^2}{2},$$ in agreement with (21).

Evidently, one may choose an opposite way, viz., starting from a linear
spectrum as the actual spectrum of a nonet, to derive the  
Gell-Mann--Okubo mass
formula. To this end, one should first calculate the average mass  
squared, Eq.
(21). Then one has to place 9 nonet states in the interval  
$(m_1,\;m_8)$ in a
way that preserves the average mass squared. As we already know,  
the isoscalar
singlet mass squared should coincide with the average mass squared;  
for the
remaining 8 states one would have the relation (16) which would in  
turn reduce
to the Gell-Mann--Okubo formula (15). One sees that the assumption  
of a linear
mass spectrum turns out to be a good alternative to group theory  
assumptions on
the form of the mass splitting, for the derivation of the  
Gell-Mann--Okubo type
relations, which may be rather difficult technical task for a  
higher symmetry
group.

We now generalize the above derivation of the Gell-Mann--Okubo mass  
formula to
the case of 4 flavors, by simply adding one more quark ($c$-quark).  
Then, in
addition to the 9 states of a nonet, we have to add 7 more states,  
to finally
form an $SU(4)$ hexadecuplet: 4 $c\bar{u},u\bar{c},c\bar{d},d\bar{c},$ 2 
$c\bar{s},s\bar{c},$ and 1 $c\bar{c}.$ Suppose that $SU(3)$  
symmetry is exact
but the underlying $SU(4)$ one is broken by the $c$-quark mass.  
Then the 9
nonet states have equal mass squared which coincide with $m_0^2,$  
and of the
remaining 7 states, 6 (which are the combinations of $c$-quark with  
one of
$u,d,s)$ have equal masses as well. Now we have to place 7 states in the 
interval $(m_0,\;m_{c\bar{c}})$ in a way that preserves the average mass
squared ($q$ stands for one of $u,d,s):$
\beq
\frac{6m_{c\bar{q}}^2+m_{c\bar{c}}^2}{7}=\frac{m_0^2+m_{c\bar{c}}^2}{2};
\eeq
therefore
\beq
12m_{c\bar{q}}^2=5m_{c\bar{c}}^2+7m_0^2,
\eeq
which is the generalization of the Gell-Mann--Okubo mass formula to  
the case of
broken $SU(4)$ but exact $SU(3)$ symmetry.

In a real world, both, $SU(4)$ and $SU(3),$ are broken, so that we  
have to
distinguish between the masses of $c\bar{u},c\bar{d},$ and  
$c\bar{s}$ states.
Introducing the standard notations, $D(c\bar{u},c\bar{d}),$  
$D_s(c\bar{s}),$
$\Psi (c\bar{c}),$ we have to modify Eq. (23), as follows:
\beq
\frac{4m_D^2+2m_{D_s}^2+m_\Psi ^2}{7}=\frac{m_0^2+m_\Psi ^2}{2};
\eeq
thus
\beq
8m_D^2+4m_{D_s}^2=5m_\Psi ^2+7m_0^2,
\eeq
which is the Gell-Mann--Okubo mass formula for an $SU(4)$  
hexadecuplet (which
has to be accompanied by those for an $SU(3)$ nonet, Eqs. (15),(16)).
Let us check this formula for the experimentally established masses  
of the
charmed mesons, for the following multiplets (which are the only  
multiplets
for which these mesons have been discovered  
experimentally)\footnote{The values
of $m_0^2$ given below are calculated from Eq. (17), which reduces,  
through
(15), to $m_0^2=(m_1^2+2m_{1/2}^2)/3.$}:

1) 1 $^1S_0$ $J^{PC}=0^{-+}$ pseudoscalar mesons, $m(D)=1.87$ GeV,  
$m(D_s)=
1.97$ GeV, $m(\eta _c)=2.98$ GeV, and $m_0^2=0.17$ GeV$^2.$  
Therefore, one
obtains 43.5 GeV$^2$ on the l.h.s. of Eq. (26), vs. 45.6 GeV$^2$ on  
the r.h.s.

2) 1 $^3S_1$ $J^{PC}=1^{--}$ vector mesons, $m(D^\ast )=2.01$ GeV,  
$m(D^\ast _
s)=2.11$ GeV, $m(J/\Psi )=3.09$ GeV, and $m_0^2=0.73$ GeV$^2.$  
Therefore, one
has 50.1 GeV$^2$ on the l.h.s. of Eq. (26) vs. 52.8 GeV$^2$ on the r.h.s.

3) 1 $^1P_1$ $J^{PC}=1^{+-}$ pseudovector mesons, $m(D_1)=2.42$  
GeV, $m(D_{s1})
=2.54$ GeV, $m(h_c(1P))=3.53$ GeV, and $m_0^2=1.58$ GeV$^2.$ In  
this case one
has 72.7 GeV$^2$ on the l.h.s of (26) vs. 73.3 GeV$^2$ on the r.h.s.

4) 1 $^3P_2$ $J^{PC}=2^{++}$ tensor mesons\footnote{Although the  
$D_{s2}^\ast
(2573)$ meson was omitted from the recent Meson Summary Table as ``needs 
confirmation'', it is the best candidate on the 1 $^3P_2$ $2^{++}$
$c\bar{s}$-state since its width and decay modes are consistent  
with $J^P=2^{
+}$ \cite{data1}.}, $m(D_2^\ast )=2.46$ GeV, $m(D_{s2}^\ast )=2.57$  
GeV, $m(
\chi _{c2}(1P))=3.55$ GeV, and $m_0^2=1.94$ GeV$^2.$ Now one has  
74.8 GeV$^2$
on the l.h.s of (26) vs. 76.5 GeV$^2$ on the r.h.s.

Thus, in either case, the relative error does not exceed 5\%, in  
the third
case the results almost coincide.

One may try to generalize the presented method of the derivation of the
Gell-Mann--Okubo formula from a linear mass spectrum to a higher  
symmetry
group. Let us briefly discuss this point. Suppose, as previously,  
that $SU(N)$
flavor symmetry is exact but the underlying $SU(N+1)$ one is broken  
by the mass
of the $(N+1)$-quark. Then we have a mass degenerate $SU(N)$  
multiplet $(N^2$
states), and $2N+1$ more states, of which $2N$ are mass degenerate  
as well.
Now we have to place these $2N+1$ states in the mass interval  
$(m_{(N,N)},\;
m_{(N+1,N+1)})$ in a way that preserves the average mass squared:
\beq
\frac{2Nm_{(N,N+1)}^2+m_{(N+1,N+1)}^2}{2N+1}=\frac{m_{(N,N)}^2+m_{(N+1,N+1)}^
2}{2}.
\eeq
This is the analog of the Gell-Mann--Okubo mass formula for an $SU(N+1)$ 
multiplet. It one writes down roughly ($m_N$ and $m_{N+1}$ are the  
masses of
$N$- and $(N+1)$-quarks, respectively)
\beq
m_{(N,N)}\simeq 2m_N,\;\;m_{(N,N+1)}\simeq  
m_N+m_{N+1},\;\;m_{(N+1,N+1)}\simeq
2m_{N+1}
\eeq
and neglects $m_N^2$ in comparison with $m_{N+1}^2,$ one obtains
\beq
m_{N+1}\simeq \frac{2N}{N-1}m_N.
\eeq
Now it becomes clear why the quark masses are exactly what they  
are. If one
goes from $SU(3)$ to $SU(4),$ one obtains from (29) $m_4\simeq  
3m_3,$ i.e.,
$m_c\simeq 3m_s,$ where $m_s$ is the constituent $s$-quark mass.  
With $m_s
\simeq 0.5-0.55$ GeV, one has $m_c\simeq 1.5-1.6$ GeV, exactly the  
value which
may be expected from the naive quark model and which is indicated by the 
recent Particle Data Group \cite{data}. If one further goes from  
$SU(4)$ to
$SU(5),$ one obtains $m_5\simeq 8/3\;m_4,$ i.e., $m_b\simeq  
8/3\;m_c.$ With
$m_c\simeq 1.5-1.6$ GeV, this implies $m_b\simeq 4.0-4.3$ GeV, again in 
agreement with the value $4.1-4.5$ GeV given by the recent Particle  
Data Group
\cite{data} and the expectations from the naive quark model. We therefore
expect the corresponding Gell-Mann--Okubo formula for an $SU(5)$  
25-plet to be
in a fair agreement with the experiment. Unfortunately, if one goes from 
$SU(5)$ to $SU(6),$ one gets the value of the top-quark mass  
$m_t\simeq 5/2
\;m_b\simeq 10-11$ GeV, which is one order of magnitude less than  
that expected
from the Standard Model\footnote{We note, however, that the low-energy 
relation between the $b$- and $t$-quark masses obtained by Jungman by 
performing a detailed numerical study of the Yukawa-coupling
renormalization-group flow in $SO(10)$ models is \cite{Jun}  
$m_b/m_t\simeq
2.5-2.7.$} \cite{data}. The explanation of this fact may lie in the mass 
spectrum deviating from a linear form for the masses of the order  
of $\sim O(
m_b).$ (E.g., the mass spectrum may be of the form $\tau (m)\sim  
m\cdot f(m),$
where $f(m)$ is close to unity for the masses up to $\sim O(m_b)$  
and begins
to grow rapidly for higher masses.) In fact, in a manifestly  
covariant theory,
the mass spectrum is $\tau (m)\sim m\cdot \exp (\alpha m^2),$ but  
normally one
expects $\alpha $ to be very small. We also note that the actual  
mass spectrum
cannot be of the Hagedorn form (6) because the exponent $a$ in Eq.  
(6) is
always negative (and is related to the number of transverse  
dimensions of a
string theory \cite{Dienes}), and therefore cannot be equal to  
unity. More
detailed discussion on this point, as well as the Gell-Mann--Okubo  
formulas
for $SU(4)$ baryon multiplets, will be given in a separate publication.

\section*{Acknowledgements}
One of us (L.B.) wish to thank E.V. Shuryak for very valuable  
discussions on
hadronic resonance spectrum.

\end{document}